\documentclass[reprint,twocolumn,groupedaddress]{revtex4-2}
\usepackage{amsmath}
\usepackage{amssymb}
\usepackage{graphicx}
\usepackage{dcolumn}
\usepackage{bm}

\begin{document}

\title{Bound States in the Continuum and Unidirectional Guided Resonances Enabled by Competing Fourier Harmonics near the Fourth Stop Band}

\author{Sun-Goo Lee}
\email{sungooleee@gmail.com}

\author{Wook-Jae Lee}
\email{wookjaelee@kongju.ac.kr}
\affiliation{Department of Data Information and Physics, Kongju National University, Gongju, 32588, South Korea}
\affiliation{Institute of Application and Fusion for Light, Kongju National University, Cheonan, 31080, South Korea}
\date{\today}

\begin{abstract}
Bound states in the continuum (BICs) and unidirectional guided resonances (UGRs) are singular radiative eigenstates in planar photonic lattices characterized by complete or directional suppression of out-of-plane radiation, respectively. Here, we show that, near the fourth stop band, the formation of BICs and UGRs can be understood within a unified two-pathway radiation model, in which the Bloch eigenmode radiates through indirect and direct pathways mediated by the first and second Fourier harmonics of the dielectric modulation, respectively. In vertically symmetric photonic lattices, when the two radiation components cancel each other out in both the upward and downward directions, BICs emerge. The model yields an analytical BIC condition that can be satisfied by tuning a single lattice parameter, the duty cycle $\rho$, and is confirmed by finite-element simulations. By contrast, in vertically asymmetric photonic lattices, the upward and downward radiation-cancellation conditions no longer coincide, allowing UGRs to arise when the cancellation condition is satisfied in only one direction. We introduce a zero-contrast grating (ZCG), in which the upper protruding grating region produces pronounced up-down asymmetry, and show that two off-$\Gamma$ UGRs emerge and can be merged at the $\Gamma$ point by tuning the lattice parameters.
\end{abstract}

\maketitle

\section{Introduction}

Subwavelength planar metastructures, including one-dimensional (1D) gratings, two-dimensional (2D) photonic crystal slabs, and metasurfaces, have become versatile platforms for controlling light--matter interactions~\cite{Chang-Hasnain2018,YHKo2018}. By tailoring the spatial distribution of the dielectric permittivity within the periodic unit cell, the radiative coupling of lattice eigenstates to the radiation continuum can be precisely engineered~\cite{SFan2002}. In particular, bound states in the continuum (BICs) and unidirectional guided resonances (UGRs) have attracted considerable attention because they support ultrahigh radiative $Q$ factors and unidirectional emission, respectively. BICs are nonradiating eigenstates embedded in the radiation continuum, for which radiation in both the upward and downward directions vanishes simultaneously~\cite{BZhen2014,Hsu2016,Minkov2018,Doeleman2018,JJin2019,MSHwang2021,MKang2023,SHan2023,Kang2025Janus,Zuo2025Janus,JYu2025}. UGRs, by contrast, arise when radiation vanishes in only one direction while remaining finite in the opposite direction, thereby enabling perfectly unidirectional light emission even in a single all-dielectric photonic-lattice slab of subwavelength thickness~\cite{Gong2023,SGLee:2024-1,Choi2025,Wang:2024,Zhuang2024,YSChoi2025,Ji2026Janus,Ding2026,Su2026}. A UGR can emerge from a BIC when in-plane $C_2$ symmetry is broken~\cite{XYin2020,SGLee2025}. Alternatively, UGRs can arise through interband coupling between distinct modes~\cite{SGLee2020-1,Yin:2023}.

In general, a guided mode can exhibit multiple stop bands associated with different orders of Bragg scattering~\cite{YDing2007}. To date, however, most experimental and theoretical studies of UGRs in planar photonic lattices have focused on regimes near the second stop band~\cite{Kazarinov1985,SGLee2019-1}. Because dielectric materials commonly used to form photonic-lattice slabs have refractive indices of approximately 2--3.5, the second stop band typically lies within the zero-order diffraction window, where higher diffraction orders are evanescent. Bloch modes in this specular regime are well suited for photonic applications because they can couple to external plane waves through the zeroth diffraction order without energy loss to higher diffraction orders~\cite{Chang-Hasnain2018,YHKo2018,SFan2002}. Far less attention has thus far been paid to the higher-order stop bands, although there appears to be no fundamental reason for this imbalance. Our recent studies have shown that Bloch modes near the fourth stop band can lie within the zero-order diffraction window and that BICs can be realized by tuning the lattice parameters~\cite{SGLee2021-2,SGLee2024-2}. An essential difference between the second and fourth stop bands lies in the radiation mechanism. Near the fourth stop band, out-of-plane radiation is mediated by the first and second Fourier harmonics of the periodic dielectric modulation, whereas near the second stop band it is mediated by the first Fourier harmonic alone~\cite{SGLee2021-1}.

Our previous studies showed that Bloch modes near the fourth stop band can support BICs and that their radiation is governed primarily by the second Fourier harmonic of the dielectric modulation, with the first Fourier harmonic providing an additional radiation pathway~\cite{SGLee2021-2,SGLee2024-2}. However, the analytical mechanism by which these two radiation pathways interfere to generate and merge BICs has not yet been fully clarified. In this study, we first develop a two-pathway radiation model that explicitly describes the destructive interference between the radiation components generated by the first and second Fourier harmonics in vertically symmetric photonic lattices. The model reveals an analytical radiation-cancellation condition for the merging of BICs at the $\Gamma$ point. We then extend the model to vertically asymmetric photonic lattices and show that UGRs can emerge when the radiation cancellation occurs in only one direction. However, although vertical asymmetry makes UGR formation possible, achieving complete one-sided radiation cancellation is generally nontrivial. We therefore introduce a zero-contrast grating (ZCG), in which the upper protruding grating region provides stronger control over the interference between the two radiation pathways. Rigorous finite-element method (FEM) simulations show that two off-$\Gamma$ UGRs emerge and can be merged at the $\Gamma$ point by tuning the lattice parameters. UGRs in the ZCG near the fourth stop band are advantageous for practical implementation because they can be realized without breaking the in-plane $C_2$ symmetry and without requiring interband coupling, allowing the use of a simpler unit-cell geometry. Our results provide a unified physical picture of BIC and UGR formation near the fourth stop band and highlight the potential of higher-order stop bands for controlling radiative eigenstates in planar photonic lattices.

\begin{figure*}[t]
\centering
\includegraphics[width=9cm]{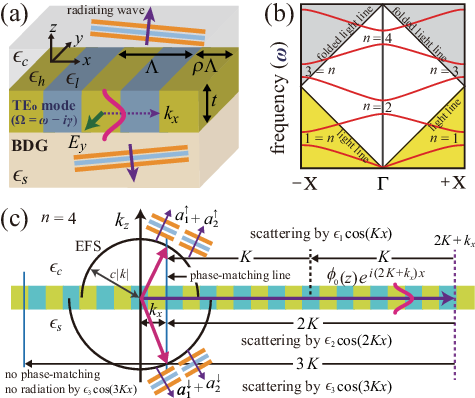}
\caption {\label{fig1} (a) Schematic of a 1D BDG. (b) Conceptual illustration of the photonic band structure of the fundamental $\mathrm{TE}_{0}$ mode and the associated Bragg stop bands. The fourth stop band ($n=4$) is located within the zero-order diffraction window (white region). (c) Phase-matching diagram on the equifrequency surface (EFS) for the fourth stop band. Out-of-plane radiation is generated by two distinct radiation pathways, $a_{1}^{\uparrow/\downarrow}$ and $a_{2}^{\uparrow/\downarrow}$, associated with the first ($q=1$) and second ($q=2$) Fourier harmonics, respectively. Higher-order harmonics ($q \ge 3$) do not satisfy the phase-matching condition, leading to no radiation. }
\end{figure*}

\section{Radiation mechanism and numerical method}

Figure~\ref{fig1}(a) illustrates a representative 1D photonic lattice, a binary diffraction grating (BDG) consisting of regions of high ($\epsilon_h$) and low ($\epsilon_l$) dielectric constant. The period and thickness of the lattice are $\Lambda$ and $t$, respectively, and the width of the high--dielectric-constant region is $\rho\Lambda$. The average dielectric constant of the lattice is kept at the constant value $\epsilon_{\mathrm{avg}}=\epsilon_l+\rho\Delta\epsilon=3.48^2$, corresponding to that of silicon at 1550 nm, even as the duty cycle $\rho$ is varied. Here, $\Delta\epsilon=\epsilon_h-\epsilon_l$ denotes the modulation depth of the dielectric constant. Figure~\ref{fig1}(b) schematically shows the photonic band structure of the fundamental TE$_0$ mode and its associated stop bands, which open at the Bragg condition $\beta=nK/2$, where $\beta$ is the propagation constant of the guided mode, $K=2\pi/\Lambda$ is the magnitude of the grating vector, and $n$ is an integer representing the Bragg order. In our analysis, the lattice parameters are chosen such that the BDG supports the fourth stop band ($n=4$) of the fundamental $\mathrm{TE}_{0}$ mode within the zero-order diffraction window (white region in Fig.~\ref{fig1}(b)), while Bloch modes near this stop band remain decoupled from higher-order guided modes, maintaining their modal purity.

To analyze the radiative properties of Bloch modes, we start with the 2D scalar wave equation for the $E_y$ field~\cite{Yariv1984}:
\begin{equation}
\left( \frac{\partial^2}{\partial x^2} + \frac{\partial^2}{\partial z^2} \right) E_y(x,z) + \epsilon(x,z)k_0^2 E_y(x,z) = 0,
\end{equation}
where $k_0$ denotes the free-space wavenumber. For a BDG with in-plane $C_2$ symmetry, the periodic dielectric function is expanded as an even cosine Fourier series, $\epsilon(x,z) = \epsilon_0(z) + \sum_{q \ge 1} \epsilon_q(z) \cos(qKx)$. Here, the Fourier coefficients are given by $\epsilon_q=(2\Delta\epsilon/q\pi)\sin(q\pi\rho)$. The coefficient $\epsilon_q$ is constant within the grating layer and zero outside the grating layer. By Bloch's theorem, the electric field can be exactly expanded in spatial harmonics as
\begin{equation}
E_y(x,z) = \sum_m E_m(z) e^{i(k_x + mK)x}.
\end{equation}
Near the fourth stop band, the dominant guided components are the $m=\pm2$ spatial harmonics, while the $m=0$ spatial harmonic represents the radiative component of the Bloch mode. Substituting these expansions into Eq. (1) yields a set of coupled ordinary differential equations for the harmonic amplitudes $E_m(z)$:
\begin{equation}
L_m E_m(z) = -\frac{k_0^2}{2} \sum_{q \ge 1} \epsilon_q(z) \left[ E_{m-q}(z) + E_{m+q}(z) \right],
\end{equation}
where $L_m \equiv d^2/dz^2 + \epsilon_0(z)k_0^2 - (k_x + mK)^2$.

Accordingly, we focus on the $m=0$ equation, which governs the radiative component of the Bloch mode:
\begin{equation}
L_0E_0(z)= -\frac{k_0^2}{2} \sum_{q \ge 1} \epsilon_q(z) \left[ E_q(z) + E_{-q}(z) \right] \equiv S(z).
\label{source}
\end{equation}
Here, $S(z)$ acts as the effective source that drives the out-of-plane radiation.

In principle, numerous Fourier harmonics ($\epsilon_q(z)\cos(qKx)$) couple different spatial field harmonics ($E_m(z)e^{i(k_x+mK)x}$) and thereby contribute to the effective radiation source. However, previous studies have shown that, near the fourth stop band, radiation loss is mediated by the first ($\epsilon_{1}\cos(Kx)$) and second ($\epsilon_{2}\cos(2Kx)$) Fourier harmonics \cite{SGLee2021-2,SGLee2024-2}. Higher-order harmonics ($\epsilon_{q \geq 3}\cos(qKx)$), which have shorter periods $p_{q \geq 3}=\Lambda/q$ and larger grating vectors $K_{q}=qK$, do not satisfy the phase-matching condition. Motivated by these considerations, we neglect the contributions from the higher-order Fourier harmonics with $q\geq3$ and approximate the effective source as
\begin{equation}
\begin{split}
S(z) \simeq -\frac{k_0^2}{2} \Big\{ &\epsilon_1(z)\left[E_1(z)+E_{-1}(z)\right]\\
&+ \epsilon_2(z)\left[E_2(z)+E_{-2}(z)\right] \Big\}.
\end{split}
\label{source-2}
\end{equation}

Retaining only the leading-order coupling in the coupled equations for $m=\pm1$, we obtain $L_{\pm1}E_{\pm1}\simeq-(k_0^2/2)\epsilon_1E_{\pm2}$, so that $E_{\pm1}\simeq-(k_0^2/2)G_{\pm1}[\epsilon_1E_{\pm2}]$. Here, $G_{\pm1}=L_{\pm1}^{-1}$ denotes the Green operator associated with the intermediate $m=\pm1$ spatial harmonic. Because the dielectric modulation $\epsilon_1(z)$ is confined to the patterned layer, $G_{\pm1}$ acts only on sources within $|z|\leq t/2$. Substituting this relation into Eq.~(\ref{source-2}) gives the general leading-order radiation source
\begin{equation}
\begin{split}
S(z)\simeq&
\frac{k_0^4}{4}\epsilon_1(z) \left\{ G_{+1}\left[\epsilon_1(z)E_2(z)\right] +G_{-1}\left[\epsilon_1(z)E_{-2}(z)\right] \right\}\\
&-\frac{k_0^2}{2}\epsilon_2(z) \left[E_2(z)+E_{-2}(z)\right].
\end{split}
\label{source-general}
\end{equation}
Equation~(\ref{source-general}) reveals two distinct radiation pathways contributing to the out-of-plane radiation. The second term represents direct coupling by the second Fourier harmonic $\epsilon_2\cos(2Kx)$, in which the $m=\pm2$ guided harmonics directly couple to the radiative $m=0$ field component. The first term represents an indirect coupling pathway mediated by the first Fourier harmonic $\epsilon_1\cos(Kx)$, in which the $m=\pm2$ harmonics first couple to the intermediate $m=\pm1$ harmonics and subsequently couple to the radiative $m=0$ field component. Because both direct and indirect pathways contribute simultaneously to the out-of-plane radiation, the amplitude of the outgoing radiative plane wave emitted by the Bloch mode is determined by the interference between the radiative components originating from the first and second Fourier harmonics. 
The radiative field driven by this source can be expressed as 
\begin{equation}
E_0(z)=\int_{-t/2}^{t/2}G(z,z')S(z')\,dz', 
\label{amplitude-general}
\end{equation}
where $G(z,z')$ is the outgoing Green function of the corresponding three-layer reference structure~\cite{Tomas1995,Santos2024}. Outside the grating layer, the radiative field takes the form of outgoing plane waves, $E_0(z)=a_{\mathrm{tot}}^{\uparrow}e^{i\mu_c(z-t/2)}$ for $z>t/2$ and $E_0(z)=a_{\mathrm{tot}}^{\downarrow}e^{-i\mu_s(z+t/2)}$ for $z<-t/2$, where $\mu_{c,s}=\sqrt{\epsilon_{c,s}k_0^2-k_x^2}$ are the vertical wavenumbers in the cover and substrate, respectively. Here, $a_{\mathrm{tot}}^{\uparrow}$ and $a_{\mathrm{tot}}^{\downarrow}$ denote the far-field radiation amplitudes in the cover and substrate, respectively, and can be written as
\begin{equation}
a_{\mathrm{tot}}^{\uparrow/\downarrow}=a_1^{\uparrow/\downarrow}+a_2^{\uparrow/\downarrow}.
\label{radiation-amplitude}
\end{equation}
The two terms $a_1^{\uparrow/\downarrow}$ and $a_2^{\uparrow/\downarrow}$ denote the radiation amplitudes associated with the indirect and direct radiation pathways, respectively.

\begin{figure*}[t]
\includegraphics[width=11cm]{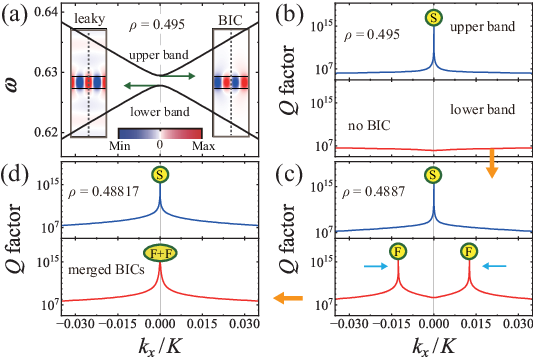}
\centering
\caption {\label{fig2} FEM results near the fourth stop band in an up-down symmetric BDG with $t=0.40~\Lambda$, $\Delta \epsilon =2.0$, and $\epsilon_s=\epsilon_c=1.46^2$. (a) Simulated dispersion relation and electric-field ($E_y$) distributions for $\rho=0.495$. The upper and lower band-edge modes are an antisymmetric symmetry-protected BIC and a symmetric radiative leaky mode, respectively. (b)--(d) Radiative $Q$ factors for $\rho=0.495$, $0.4887$, and $0.48817$, respectively. In (b), no Fourier BIC is observed in the lower band. In (c), two Fourier BICs, denoted by F, are located in the lower band at $k_x/K=\pm0.0126$. In (d), the two Fourier BICs merge at the $\Gamma$ point, as denoted by F+F.}
\end{figure*}

In general, evaluating the radiation amplitudes requires detailed knowledge of the field profiles $E_{\pm2}(z)$ of the spatial harmonics of the Bloch mode. At the $\Gamma$ point, however, the two band-edge modes satisfy the symmetry relations $E_2(z)=E_{-2}(z)$ or $E_2(z)=-E_{-2}(z)$, while the Green operators associated with the $m=\pm1$ spatial harmonics become identical, $G_{+1}=G_{-1}\equiv G_1$. These relations greatly simplify the radiation amplitudes and provide useful insight into the physical conditions governing the formation of BICs and UGRs near the fourth stop band. To obtain explicit expressions at the third-order $\Gamma$ point (corresponding to the fourth-order Bragg condition), we further approximate the $m=\pm2$ guided harmonics as $E_2(z)\simeq A\phi(z)$ and $E_{-2}(z)\simeq B\phi(z)$, where $\phi(z)$ denotes the transverse profile of the fundamental $\mathrm{TE}_0$ mode of the unmodulated slab, and $A$ and $B$ are the corresponding complex modal coefficients. Because $\epsilon_{\mathrm{avg}}$ is kept fixed, $\phi(z)$ remains unchanged and can be used as a fixed reference basis as $\rho$ is varied. The indirect contributions are then given by
\begin{align}
a_1^{\uparrow} &= \frac{k_0^4\tau_c e^{i\mu t/2}}{4D}(A+B)\epsilon_1^2 J_1^{\uparrow}, \label{a1-up}\\
a_1^{\downarrow} &= \frac{k_0^4\tau_s e^{i\mu t/2}}{4D}(A+B)\epsilon_1^2 J_1^{\downarrow}, \label{a1-down}
\end{align}
and the direct contributions are given by
\begin{align}
a_2^{\uparrow} &= -\frac{k_0^2\tau_c e^{i\mu t/2}}{2D}(A+B)\epsilon_2 J_2^{\uparrow}, \label{a2-up}\\
a_2^{\downarrow} &= -\frac{k_0^2\tau_s e^{i\mu t/2}}{2D}(A+B)\epsilon_2 J_2^{\downarrow}. \label{a2-down}
\end{align}
Here, the overlap integrals $J_1^{\uparrow/\downarrow}$ and $J_2^{\uparrow/\downarrow}$ associated with the indirect and direct radiation pathways, respectively, are defined as
\begin{align}
J_1^{\uparrow} &= \int_{-t/2}^{t/2}G_1[\phi(z)]
\left[e^{-i\mu z}+r_s e^{i\mu t}e^{i\mu z}\right]dz, \label{J1-up}\\
J_1^{\downarrow} &= \int_{-t/2}^{t/2}G_1[\phi(z)]
\left[e^{i\mu z}+r_c e^{i\mu t}e^{-i\mu z}\right]dz, \label{J1-down}\\
J_2^{\uparrow} &= \int_{-t/2}^{t/2}\phi(z)
\left[e^{-i\mu z}+r_s e^{i\mu t}e^{i\mu z}\right]dz, \label{J2-up}\\
J_2^{\downarrow} &= \int_{-t/2}^{t/2}\phi(z)
\left[e^{i\mu z}+r_c e^{i\mu t}e^{-i\mu z}\right]dz. \label{J2-down}
\end{align}
In Eqs.~(\ref{J1-up})--(\ref{J2-down}), $\mu=\sqrt{\epsilon_{\mathrm{avg}}k_0^2-k_x^2}$ is the vertical wavenumber in the grating layer. The coefficients $r_c$ and $r_s$ denote the reflection coefficients at the grating--cover and grating--substrate interfaces, respectively, while $\tau_c$ and $\tau_s$ denote the corresponding transmission coefficients, with $r_{c,s}=(\mu-\mu_{c,s})/(\mu+\mu_{c,s})$ and $\tau_{c,s}=2\mu/(\mu+\mu_{c,s})$. The common denominator $D=2i\mu(1-r_cr_se^{2i\mu t})$ accounts for multiple reflections within the grating layer.

According to Eqs.~(\ref{a1-up})--(\ref{J2-down}), complete cancellation of out-of-plane radiation in either the upward or downward direction, $a_1^{\uparrow/\downarrow}+a_2^{\uparrow/\downarrow}=0$, is achieved when the following condition is satisfied:
\begin{equation}
\frac{\epsilon_2 J_2^{\uparrow/\downarrow}}{\epsilon_1^2 J_1^{\uparrow/\downarrow}}=\frac{k_0^2}{2}.
\label{overlap-condition}
\end{equation}
In vertically symmetric lattices, the upward and downward radiation amplitudes are identical, with $J_q^{\uparrow}=J_q^{\downarrow}$. A BIC can emerge when Eq.~(\ref{overlap-condition}) is satisfied simultaneously for both the upward and downward directions. We refer to such BICs as Fourier BICs because the radiation vanishes due to destructive interference between the radiation components generated by different Fourier harmonics. In vertically asymmetric lattices, however, $J_q^{\uparrow}\neq J_q^{\downarrow}$ in general, and the upward and downward radiation amplitudes are therefore unequal. Consequently, the cancellation conditions for the upward and downward radiation no longer coincide. Nevertheless, the total radiation amplitude can still vanish in one of the two directions, such that $a_{\mathrm{tot}}^{\uparrow}=0$ while $a_{\mathrm{tot}}^{\downarrow}\neq0$, or vice versa. This one-sided radiation cancellation constitutes a UGR. Equations~(\ref{a1-up})--(\ref{a2-down}) also indicate that all radiation amplitudes vanish when $A+B=0$. In photonic lattices with in-plane $C_2$ symmetry, it is well established that one of the band-edge modes has an antisymmetric field distribution along the $x$ direction, for which $A+B=0$, yielding a symmetry-protected BIC irrespective of vertical mirror symmetry.

The analytical model is evaluated as follows. For a given slab thickness $t$ and cladding indices, the transverse profile $\phi(z)$ and the resonance wavenumber $k_0$ of the TE$_0$ mode at $\beta=2K$ are obtained from the three-layer dispersion relation. The intermediate field $u(z)\equiv G_1[\phi(z)]$ is then obtained by solving $L_1u(z)=\phi(z)$ on a uniform grid using finite differences, with the source term restricted to the grating layer and the grid chosen such that the layer boundaries $z=\pm t/2$ coincide with grid points. Because the $m=\pm1$ harmonic is evanescent outside the grating layer, $u(z)$ was set to zero at the edges of a computational window extending far beyond the evanescent tails of the mode. The overlap integrals in Eqs.~(\ref{J1-up})--(\ref{J2-down}) are subsequently evaluated by numerical quadrature over the grating layer. The FEM results were obtained using the eigenfrequency solver in COMSOL Multiphysics, with a physics-controlled mesh and the element size set to Extremely fine. A computational unit cell of size $\Lambda \times 12\Lambda$ along the $x$ and $z$ directions, respectively, was discretized, with Floquet periodic boundary conditions imposed along $x$ and perfectly matched layers terminating the cover and substrate along $z$. The resulting complex eigenfrequencies $\Omega=\omega-i\gamma$ of the TE Bloch modes were used to evaluate the radiative quality factor $Q=\omega/(2\gamma)$. For the analytical model, the radiation ratio was evaluated from the far-field radiation amplitudes as $\eta=\mu_c|a_{\mathrm{tot}}^{\uparrow}|^2/(\mu_s|a_{\mathrm{tot}}^{\downarrow}|^2)$. For the FEM simulations, the upward and downward radiated powers $P_{\mathrm{up}}$ and $P_{\mathrm{down}}$ were obtained by integrating the time-averaged Poynting flux through planes in the cover and substrate, respectively, and the radiation ratio was evaluated as $\eta=P_{\mathrm{up}}/P_{\mathrm{down}}$. In both cases, $\eta$ was presented in decibels as $10\log_{10}\eta$.

\section{Analytical verification of Fourier BICs}

Fourier BICs near the fourth stop band have previously been identified through numerical simulations \cite{SGLee2021-2,SGLee2024-2}. However, a clear analytical understanding of their physical origin has remained lacking. Here, using the two-pathway radiation model developed in the preceding section, we show that these BICs arise from destructive interference between the radiation components generated by the direct and indirect pathways. Figure~\ref{fig2}(a) shows the FEM-simulated dispersion relation of an up-down symmetric BDG with $t=0.40~\Lambda$, $\Delta \epsilon =2.0$, $\epsilon_s=\epsilon_c=1.46^2$, and $\rho=0.495$. The Bloch modes near the fourth stop band are well isolated from other guided modes. The spatial electric-field ($E_y$) distributions shown in the insets indicate that the upper band-edge mode is an antisymmetric symmetry-protected BIC, whereas the lower band-edge mode is a symmetric radiative leaky mode, with both modes originating from the fundamental $\mathrm{TE}_0$ mode. Their $\Lambda/2$ periodicity further indicates that both modes are located at the third-order $\Gamma$ point corresponding to the fourth-order Bragg condition ($n=4$). Figures~\ref{fig2}(b)--\ref{fig2}(d) show the radiative $Q$ factors for three different duty cycles, $\rho=0.495$, $0.4887$, and $0.48817$, respectively, at fixed $t=0.40~\Lambda$. For all three duty cycles, the divergence of the radiative $Q$ factor at the upper band edge confirms the presence of a symmetry-protected BIC, denoted by S, at the $\Gamma$ point. At $\rho=0.495$, no BIC is observed in the lower band [Fig.~\ref{fig2}(b)]. When $\rho$ is reduced to $0.4887$, two Fourier BICs, denoted by F, emerge in the lower band at $k_x/K=\pm0.0126$ [Fig.~\ref{fig2}(c)]. As $\rho$ is further reduced, these two Fourier BICs move toward the $\Gamma$ point and eventually merge at $k_x=0$ for $\rho=0.48817$ [Fig.~\ref{fig2}(d)].

\begin{figure}[]
\includegraphics[width=7cm]{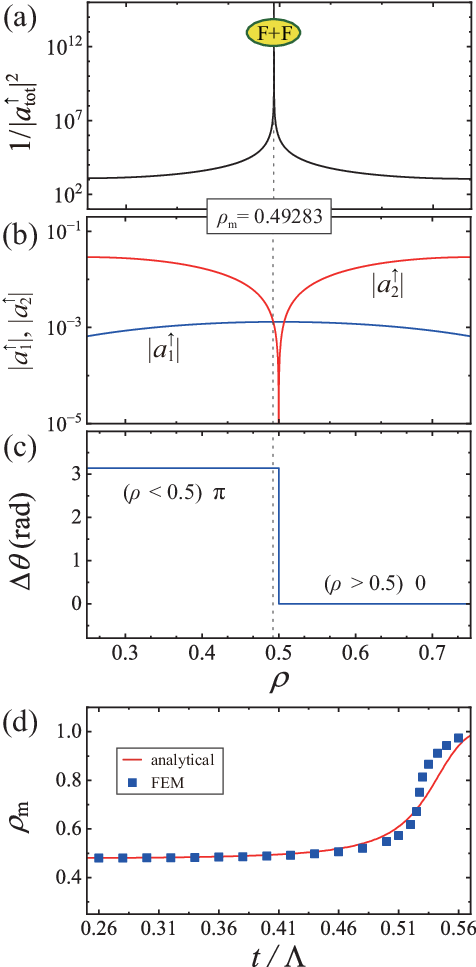}
\centering
\caption {\label{fig3} (a) Inverse squared magnitude of the total upward radiation amplitude, $1/|a_{\mathrm{tot}}^{\uparrow}|^2$, at the $\Gamma$ point as a function of the duty cycle $\rho$ for $t=0.40~\Lambda$. The divergence at $\rho=\rho_{\mathrm{m}}=0.49283$ indicates the merging condition of the two Fourier BICs, denoted by F+F. (b) Magnitudes of the two radiation amplitudes, $|a_1^{\uparrow}|$ and $|a_2^{\uparrow}|$, showing equal amplitudes at $\rho=\rho_{\mathrm{m}}$. (c) Phase difference $\Delta\theta=\arg(a_1^{\uparrow})-\arg(a_2^{\uparrow})$, showing $\Delta\theta=\pi$ for $\rho<0.5$ and $\Delta\theta=0$ for $\rho>0.5$. (d) Merging duty cycle $\rho_{\mathrm{m}}$ as a function of the normalized slab thickness $t/\Lambda$, obtained from the analytical model and FEM simulations.}
\end{figure}

\begin{figure*}[t]
\includegraphics[width=\textwidth]{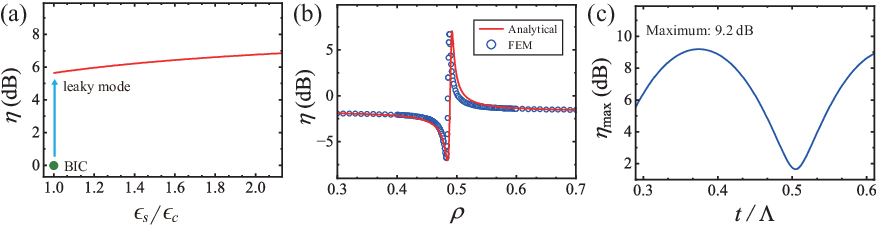}
\caption {\label{fig4} Radiation asymmetry in vertically asymmetric BDGs. (a) Radiation ratio $\eta$ calculated using the analytical model as a function of the dielectric-constant ratio $\epsilon_s/\epsilon_c$ at $\rho=\rho_m=0.49283$ and $t=0.40\Lambda$. (b) Radiation ratio $\eta$ as a function of the duty cycle $\rho$ for an air-cover/silica-substrate BDG with $t=0.40\Lambda$, obtained from the analytical model (red solid line) and FEM simulations (blue circles). (c) Maximum radiation ratio $\eta_{\max}$ as a function of the normalized slab thickness $t/\Lambda$, obtained by optimizing $\rho$ at each thickness.}
\end{figure*}

Previous studies reported the merging of accidental BICs at the second stop band as the slab thickness was varied \cite{SGLee2023}. In Figs.~\ref{fig2}(b)--\ref{fig2}(d), however, tuning $\rho$ continuously changes the balance between the two radiation pathways, causing the Fourier BICs to shift toward and eventually merge at the $\Gamma$ point. This behavior is fundamentally different from the evolution of accidental BICs reported at the second stop band. Figure~\ref{fig3}(a) shows the inverse squared magnitude of the total radiation amplitude, $1/|a_{\mathrm{tot}}^{\uparrow}|^2$, at the $\Gamma$ point as a function of $\rho$, calculated using the two-pathway radiation model for the same up-down symmetric BDG considered in Fig.~\ref{fig2}. Since the BDG is up-down symmetric, $a_{\mathrm{tot}}^{\uparrow}=a_{\mathrm{tot}}^{\downarrow}$, $a_1^{\uparrow}=a_1^{\downarrow}$, and $a_2^{\uparrow}=a_2^{\downarrow}$. A sharp divergence occurs at $\rho=\rho_{\mathrm{m}}=0.49283$, indicating complete cancellation of the total radiation amplitude at the $\Gamma$ point. Because the analytical model is formulated at the $\Gamma$ point, it does not explicitly track the trajectories of the off-$\Gamma$ Fourier BICs. However, in a lattice preserving in-plane $C_2$ symmetry, if a Fourier BIC occurs at $+k_x$, a corresponding Fourier BIC simultaneously occurs at $-k_x$. Therefore, when the Fourier-BIC cancellation condition is satisfied at $k_x=0$, the two Fourier BICs coincide at the $\Gamma$ point. The duty cycle $\rho_{\mathrm{m}}$ predicted by the analytical model can thus be interpreted as the merging condition of the two Fourier BICs, as denoted by F+F in Fig.~\ref{fig3}(a). To clarify the origin of this radiation cancellation, Fig.~\ref{fig3}(b) shows the magnitudes of the two radiation amplitudes, $|a_1^{\uparrow}|$ and $|a_2^{\uparrow}|$. At $\rho=\rho_{\mathrm{m}}$, the two components have equal magnitudes, $|a_1^{\uparrow}|=|a_2^{\uparrow}|$. Figure~\ref{fig3}(c) further shows their phase difference, $\Delta\theta=\arg(a_1^{\uparrow})-\arg(a_2^{\uparrow})$. For $\rho<0.5$, the two radiation components are out of phase with $\Delta\theta=\pi$, whereas for $\rho>0.5$ they are in phase with $\Delta\theta=0$. Figures~\ref{fig3}(a)--\ref{fig3}(c) show that, at $\rho=\rho_{\mathrm{m}}=0.49283$, the simultaneous conditions $|a_1^{\uparrow}|=|a_2^{\uparrow}|$ and $\Delta\theta=\pi$ lead to complete destructive interference, $a_{\mathrm{tot}}^{\uparrow}=a_1^{\uparrow}+a_2^{\uparrow}=0$, thereby producing a Fourier BIC at the $\Gamma$ point.

Equations~(\ref{a1-up})--(\ref{a2-down}) show that the radiation amplitudes are governed by overlap integrals between the effective radiation-source distributions associated with the guided mode and the propagating plane-wave components in the grating layer. Because these integrals depend on the slab thickness, the Fourier-BIC merging condition is also expected to vary with $t/\Lambda$. Figure~\ref{fig3}(d) compares $\rho_{\mathrm{m}}$ predicted by the analytical model (red solid line) with that obtained from FEM simulations (blue solid squares). The analytical results closely follow the FEM results over $0.25\leq t/\Lambda\leq0.57$. The physical origin of this thickness dependence can be understood from the different responses of the two radiation pathways. For the up-down symmetric BDG, the relevant vertical field distributions are even functions of $z$, so the sine components in Eqs.~(\ref{J1-up})--(\ref{J2-down}) vanish. The overlap integral can be simply written as $J_1=J_1^{\uparrow/\downarrow}=\int_{-t/2}^{t/2}G_1[\phi(z)]\cos(\mu z)\,dz$ and $J_2=J_2^{\uparrow/\downarrow}=\int_{-t/2}^{t/2}\phi(z)\cos(\mu z)\,dz$ for the indirect and direct pathways, respectively. The Fourier-BIC condition $a_1^{\uparrow/\downarrow}+a_2^{\uparrow/\downarrow}=0$ then gives
\begin{equation}
\cot(\pi\rho_{\mathrm{m}})=\frac{k_0^2\Delta\epsilon}{\pi}\frac{J_1}{J_2}.
\label{merging-condition}
\end{equation}
This expression directly relates the merging duty cycle to the ratio of the two overlap factors. Since $\epsilon_1=(2\Delta\epsilon/\pi)\sin(\pi\rho)$ remains positive over $0<\rho<1$, whereas $\epsilon_2=(\Delta\epsilon/\pi)\sin(2\pi\rho)$ changes sign at $\rho=0.5$, the sign of the direct contribution depends on the sign of $\epsilon_2J_2$. In the investigated thickness range $0.25\leq t/\Lambda\leq0.57$, $J_2$ remains positive, so the sign of the direct contribution is controlled by the duty cycle alone. In contrast, $J_1$ crosses zero and changes sign from positive to negative at $t=0.42536~\Lambda$. This sign reversal corresponds to a phase change of $\pi$ in the indirect radiation amplitude $a_1$. For $J_1>0$, destructive interference between $a_1$ and $a_2$ occurs on the $\rho<0.5$ side, where $\epsilon_2>0$. At $J_1=0$, the above relation gives $\rho_{\mathrm{m}}=0.5$. Once $J_1$ becomes negative, the phase of $a_1$ is reversed, and destructive interference requires $\epsilon_2<0$, which occurs for $\rho>0.5$. Consequently, $\rho_{\mathrm{m}}$ shifts above 0.5. As the slab thickness increases further, $|J_1|$ increases while $J_2$ decreases, making $J_1/J_2$ increasingly negative. Because $\cot(\pi\rho_{\mathrm{m}})$ decreases from zero toward negative infinity as $\rho_{\mathrm{m}}$ increases from 0.5 toward 1, this trend drives $\rho_{\mathrm{m}}$ progressively toward unity. This accounts for the rapid increase of $\rho_{\mathrm{m}}$ at larger $t/\Lambda$ in Fig.~\ref{fig3}(d).

\section{Radiation asymmetry induced by broken up-down mirror symmetry}

According to the overlap integrals in Eqs.~(\ref{J1-up})--(\ref{J2-down}), the out-of-plane radiation in the indirect and direct pathways is generated by effective source distributions proportional to $G_1[\phi(z)]$ and $\phi(z)$, respectively. In a vertically symmetric structure, the TE$_0$ mode profile $ \phi(z) $ is an even function of $z$, and $G_1[\phi(z)]$ is also even. Consequently, the overlap integrals take real values, and $J_1^{\uparrow}=J_1^{\downarrow}$ and $J_2^{\uparrow}=J_2^{\downarrow}$ are simultaneously satisfied, allowing the radiation-cancellation condition in Eq.~(\ref{overlap-condition}) to be satisfied simultaneously in the upward and downward directions and thereby giving rise to a Fourier BIC. When the up-down mirror symmetry is broken, however, $\phi(z)$ and $G_1[\phi(z)]$ no longer possess even symmetry, and the overlap integrals generally become complex, with $J_1^{\uparrow}\neq J_1^{\downarrow}$ and $J_2^{\uparrow}\neq J_2^{\downarrow}$. Consequently, the radiation-cancellation conditions for the upward and downward directions no longer coincide, and a Fourier BIC is expected to transform into a radiative state with asymmetric upward and downward emission. 

A simple and direct way to break the up-down mirror symmetry in a BDG structure is to use different dielectric constants for the cover and substrate, i.e., $\epsilon_c\neq\epsilon_s$. Figure~\ref{fig4}(a) shows the radiation ratio $\eta$ calculated using the analytical model as a function of the dielectric-constant ratio $\epsilon_s/\epsilon_c$, with the duty cycle fixed at $\rho=\rho_{\mathrm{m}}=0.49283$, corresponding to the Fourier-BIC condition of the up-down symmetric BDG in Fig.~\ref{fig3}(a). At $\epsilon_s/\epsilon_c=1$, the upward and downward radiation amplitudes vanish simultaneously, forming a Fourier BIC. As $\epsilon_s/\epsilon_c$ increases above unity, the up-down mirror symmetry is broken and the Fourier BIC is converted into a leaky mode with asymmetric radiation. Even for a very small vertical asymmetry of $\epsilon_s/\epsilon_c=1.001$, the radiation ratio reaches approximately $5.6~\mathrm{dB}$. As the dielectric contrast increases further, $\eta$ increases monotonically and reaches approximately $6.8~\mathrm{dB}$ at $\epsilon_s/\epsilon_c=2.1316$, corresponding to an air cover and a silica substrate. This air/silica configuration represents the largest practically accessible vertical dielectric contrast for conventional photonic structures employing a silica substrate. For the air-cover/silica-substrate configuration, the upward and downward radiation-cancellation conditions are therefore expected to occur at duty cycles different from the symmetric-structure value of $\rho_{\mathrm{m}}=0.49283$. We therefore varied $\rho$ while fixing $t=0.40~\Lambda$ and calculated $\eta$ using the analytical two-pathway radiation model. For comparison, FEM simulations were also performed for the same structure. As shown by the red solid line in Fig.~\ref{fig4}(b), the analytical calculation yields a maximum $\eta$ of approximately $7.0~\mathrm{dB}$ at $\rho=0.4922$ and a minimum of approximately $-7.0~\mathrm{dB}$ at $\rho=0.4846$. The FEM results, shown by the blue circles, closely follow the analytical curve, demonstrating that the two-pathway radiation model provides a good description of the radiative behavior at the third-order $\Gamma$ point not only in vertically symmetric BDGs but also in vertically asymmetric BDGs.

\begin{figure*}[]
\includegraphics[width=\textwidth]{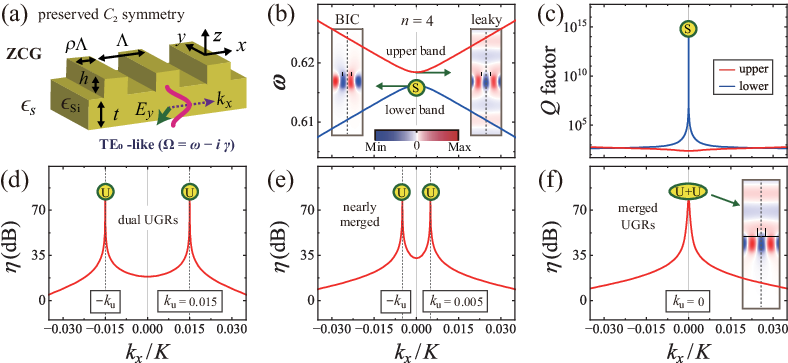}
\caption {\label{fig5} (a) Schematic of the silicon ZCG with broken up-down mirror symmetry while preserving in-plane $C_2$ symmetry. FEM-simulated (b) photonic band structure and (c) radiative $Q$ factor near the fourth stop band for the ZCG with $t=0.45~\Lambda$, $h=0.19~\Lambda$, and $\rho=0.15$. The $\mathrm{TE}_{0}$-like mode remains well separated from other guided modes, and a symmetry-protected BIC is located at the lower band edge. (d) FEM-simulated $\eta$ curve for the upper band at $t=0.45~\Lambda$, $h=0.2494~\Lambda$, and $\rho=0.2719$, showing two UGRs at $k_x/K=\pm k_{\mathrm{u}}=\pm 0.015$. (e) For $h=0.2675~\Lambda$ and $\rho=0.2303$, the dual UGRs move toward the $\Gamma$ point and are located at $k_x/K=\pm 0.005$. (f) For $h=0.2744~\Lambda$ and $\rho=0.2165$, the two UGRs merge at the $\Gamma$ point, forming a $\Gamma$-point UGR.}
\end{figure*}

As shown in Figs.~\ref{fig4}(a) and \ref{fig4}(b), the attainable radiation asymmetry remains moderate even when the vertical dielectric contrast and the duty cycle $\rho$ are varied. We therefore further varied the slab thickness $t$ and, for each thickness, scanned $\rho$ to determine the maximum attainable radiation ratio $\eta_{\max}$. Figure~\ref{fig4}(c) shows that $\eta_{\max}$ varies with $t/\Lambda$ but does not exceed $9.2~\mathrm{dB}$ over the investigated thickness range $0.25\leq t/\Lambda\leq0.57$. The origin of this limitation can be understood from the fact that in a vertically asymmetric BDG, the overlap integrals are generally complex. Complete radiation cancellation in the upward or downward direction requires the simultaneous satisfaction of the magnitude and phase conditions, i.e., $|a_1^{\uparrow/\downarrow}|=|a_2^{\uparrow/\downarrow}|$ and $\arg(a_1^{\uparrow/\downarrow})-\arg(a_2^{\uparrow/\downarrow})=\pi$. Equivalently, in terms of the overlap integrals, these conditions can be written as $(k_0^2/2)\epsilon_1^2|J_1^{\uparrow/\downarrow}|=|\epsilon_2J_2^{\uparrow/\downarrow}|$ and $\arg(J_1^{\uparrow/\downarrow})-\arg(\epsilon_2J_2^{\uparrow/\downarrow})=0$. The direct contribution $|\epsilon_2J_2^{\uparrow/\downarrow}|$ is generally larger than the indirect contribution $(k_0^2/2)\epsilon_1^2|J_1^{\uparrow/\downarrow}|$. However, because $\epsilon_2$ vanishes at $\rho=0.5$, the magnitude condition can be satisfied by varying the duty cycle $\rho$. The phase condition, however, cannot be controlled by $\rho$, because $\epsilon_2$ is real and only changes sign as $\rho$ passes through 0.5. Therefore, the single tuning parameter $\rho$ is insufficient to satisfy the magnitude and phase conditions simultaneously. In a vertically symmetric structure, however, the contributions from the two radiation pathways are real-valued, so they are either in phase ($\Delta\theta=0$) or out of phase ($\Delta\theta=\pi$). Since $\epsilon_2$ changes sign from positive for $\rho<0.5$ to negative for $\rho>0.5$, if the two contributions are in phase (out of phase) for $\rho<0.5$, they become out of phase (in phase) for $\rho>0.5$. Since the phase condition is therefore automatically satisfied on one side of $\rho=0.5$, varying the single parameter $\rho$ is sufficient to satisfy the magnitude condition and realize a BIC. 

In a vertically asymmetric BDG, changing the slab thickness $t$ modifies both the magnitudes and phases of the overlap integrals $J_1^{\uparrow/\downarrow}$ and $J_2^{\uparrow/\downarrow}$, providing an additional degree of freedom beyond the duty cycle $\rho$. Nevertheless, as evidenced by Fig.~\ref{fig4}(c), this additional tuning does not produce sufficiently high radiation asymmetry to realize a UGR over the investigated parameter range. In this structure, the vertical asymmetry is introduced only through the dielectric contrast between the cover and substrate, which only weakly modifies the vertical distributions of the effective radiation sources $\phi(z)$ and $G_1[\phi(z)]$. Consequently, the relative amplitudes and phases of the two radiation pathways cannot be controlled sufficiently to produce highly asymmetric radiation. To achieve complete one-sided radiation cancellation, stronger control over the effective radiation-source distributions is therefore required. This motivates the introduction of a ZCG \cite{Magnusson2014,MNiraula2015}, in which the vertically protruding grating region directly modifies the effective radiation-source distributions and thereby provides stronger control over the relative amplitudes and phases of the two radiation pathways.

\section{Formation of UGRs in a ZCG}

Figure~\ref{fig5}(a) shows the schematic of a silicon zero-contrast grating (ZCG) structure with broken up-down mirror symmetry while preserving the in-plane $C_2$ symmetry. Although the upper and lower surrounding media are made of the same low-index material with $\epsilon_c=\epsilon_s=1.46^2$, the upper protrusion, defined by width $\rho\Lambda$ and height $h$, breaks the up-down mirror symmetry and substantially modifies both the eigenmode field distribution and the spatial distribution of the radiation sources, providing stronger control over the interference between the two radiation pathways. Because the analytical expressions in Eqs.~(\ref{amplitude-general})--(\ref{J2-down}) are derived for the three-layer BDG structure and are not directly applicable to the ZCG geometry, the following results are obtained using rigorous FEM simulations. The simulated dispersion curves and radiative $Q$ factors plotted in Figs.~\ref{fig5}(b) and \ref{fig5}(c) show that, with the ZCG parameters $t=0.45~\Lambda$, $h=0.19~\Lambda$, and $\rho=0.15$, near the fourth stop band the $\mathrm{TE}_{0}$-like mode remains well separated from other guided modes and retains its modal purity, while a symmetry-protected BIC is formed at the lower band edge.

We performed rigorous FEM simulations over various combinations of $(t,h,\rho)$, and Figs.~\ref{fig5}(d)--\ref{fig5}(f) present the calculated $\eta$ curves for three representative cases in which UGRs emerge in the upper band for a fixed $t=0.45~\Lambda$ and different combinations of $(h,\rho)$. When $\rho=0.2719$ and $h=0.2494~\Lambda$, two UGRs with $\eta = 81~\mathrm{dB}$ appear symmetrically at $k_x/K=\pm k_{\mathrm{u}}=\pm 0.015$, as shown in Fig.~\ref{fig5}(d), owing to the preserved in-plane $C_2$ symmetry. When $\rho=0.2303$ and $h=0.2675~\Lambda$, they move toward the $\Gamma$ point and are located at $k_x/K=\pm0.005$, as shown in Fig.~\ref{fig5}(e). Upon further tuning of $\rho$ and $h$ with $t=0.45~\Lambda$ fixed, the two UGRs merge at the $\Gamma$ point for $\rho=0.2165$ and $h=0.2744~\Lambda$, as shown in Fig.~\ref{fig5}(f). These results show that the UGR positions can be controlled by tuning the lattice parameters $\rho$ and $h$ while preserving the in-plane $C_2$ symmetry. The inset in Fig.~\ref{fig5}(f) shows that the merged UGR at the $\Gamma$ point radiates along the upward surface-normal direction. The realization of a UGR at the $\Gamma$ point is practically important because it enables direct coupling to normally propagating free-space modes, making it attractive for applications such as surface-emitting lasers, optical sensors, and directional emitters. Previous studies have realized UGRs either by breaking the in-plane $C_2$ symmetry \cite{Wang:2024} or through interband coupling between distinct modes near a modal crossing \cite{Yin:2023,SGLee:2024-1}. By contrast, our mechanism exploits two radiation pathways associated with the first and second Fourier harmonics, thereby enabling the two off-$\Gamma$ UGRs to merge at the $\Gamma$ point while preserving in-plane $C_2$ symmetry and without requiring interband coupling.

In this study, the radiative $Q$ factors and radiation asymmetry are calculated for ideal lossless structures. Material loss and fabrication imperfections are expected to reduce the experimentally attainable $Q$ factors and may also perturb the radiation-cancellation condition for BICs and UGRs. A quantitative assessment of these nonideal effects requires a dedicated analysis and may be a subject of future study. The UGRs realized in the ZCG near the fourth stop band offer practical advantages. For a given operating frequency, the fourth stop band allows a lattice period approximately twice that required at the second stop band, resulting in larger structural dimensions and relaxed lithographic constraints.

\section{Conclusion}

In conclusion, we have shown that singular radiative eigenstates near the fourth stop band of planar photonic lattices are governed by a two-pathway radiation-cancellation mechanism arising from the first and second Fourier harmonics of the dielectric modulation. In up-down symmetric lattices, this mechanism generates off-$\Gamma$ Fourier BICs through destructive interference between the two radiation pathways, while breaking the up-down symmetry converts these states into directionally asymmetric resonances. The two-pathway radiation mechanism is confirmed by rigorous FEM simulations. We demonstrated that off-$\Gamma$ UGRs can be realized in ZCG structures without requiring in-plane $C_{2}$ symmetry breaking or interband coupling, and that these states can merge at the $\Gamma$ point to radiate exactly along the surface normal. Notably, the fourth stop band allows for the implementation of these singular radiative eigenstates with a lattice period approximately twice that required at the second stop band for a given frequency. This geometric advantage, combined with the preservation of in-plane symmetry, relaxes lithographic constraints and simplifies device design. Our results provide useful insight into future theoretical studies and the practical implementation of BICs and UGRs in planar photonic lattices.

\begin{acknowledgments}
This research was supported by a grant from the National Research Foundation of Korea funded by the Ministry of Science and ICT (2022R1A2C1011091).
\end{acknowledgments}

\end{document}